\documentstyle{l-aa}
\begin{document}

  \thesaurus{  11          % Galaxies
              (11.01.1;    % Galaxies:abundances
               11.05.2;    % Galaxies:evolution
               11.09.4;    % Galaxies:ISM
               11.09.5) }  % Galaxies:irregular

   \title{The oxygen abundance deficiency in irregular galaxies}

   \author{ L.S. Pilyugin \inst{1},  F.Ferrini \inst{2},}

  \offprints{L.S. Pilyugin }

   \institute{   Main Astronomical Observatory
                 of National Academy of Sciences of Ukraine,
                 Goloseevo, 03680 Kiev-127, Ukraine, \\
                 (pilyugin@mao.kiev.ua)
                  \and
                 Department of Physics, Section of Astronomy,
                 University of Pisa, piazza Torricelli 2,
                 56100 Pisa, Italy,  \\
                 (federico@astr2pi.difi.unipi.it)}
 
   \date{Received ; accepted }

\maketitle

\markboth {L.S.Pilyugin, F.Ferrini: oxygen abundance deficiency in
                irregular galaxies}{}

\begin{abstract}

The observed oxygen abundances in a number of irregular galaxies have been 
compared with predictions of the closed-box model of chemical and photometric
evolution of  galaxies. Oxygen is found to be deficient with respect to
the predicted abundances. This is an indicator in favor of loss of  heavy 
elements via galactic winds or/and of infall of low--abundance gas onto the galaxy.

The oxygen abundance deficiency observed within the optical edge of a galaxy 
cannot be explained by mixing with the gas envelope
observed outside the optical limit. We confirm the widespread idea 
that a significant part of the heavy elements is ejected by irregular
galaxies in the intergalactic medium.


\end{abstract}

\section{Introduction}

Irregular galaxies gained a central position in  extragalactic studies, as
the most abundant and least evolved galaxies in the universe.
It was soon understood that they are not closed systems, and that
galactic winds had to play a relevant role in their chemical evolution.

The hydrodynamic simulations of irregular galaxies lead to conflicting
conclusions: de Young \& Gallagher (1990), de Young \& Heckman (1994) and 
MacLow \& Ferrara (1999) found that irregulars can lose
a significant fraction of metals via galactic winds; conversely, Silich 
\& Tenorio-Tagle (1998) concluded that a low-density gaseous halo 
can inhibit the loss of matter from dwarf galaxies. Models of chemical 
evolution of irregular galaxies suggested that galactic winds can explain 
the  chemical properties of these objects (Matteucci \& Chiosi 1983; 
Matteucci \& Tosi 1985; Pilyugin 1993, 1996; Marconi et al 1994; 
Bradamante et al 1998, and many others). Outflows of ionized gas have 
been observed in a number of irregular galaxies (Marlowe et al 1995).  
Skillman (1997) has summarized the pros and cons for galactic wind--dominated 
evolution of irregular galaxies and has concluded that neither case is very strong.

%++++++++++++++++++++++++++++++++++++++++++++++++++++++ Table obsdat
\begin{table*}
\caption[]{\label{table:obsdat}
Observational characteristics (with references) for the
 galaxies included in the present sample}
\begin{flushleft}
\begin{tabular}{llcllcclcl} \hline \hline
name       &     other &T type & distance & ref      &log$L_{B}$&log$M_{HI}$ & B-V  & 12+log O/H  &  ref    \\
           &     name  &       &   Mpc    &          &         &             &      &            &         \\  \hline
 WLM       & DDO 221   &  10   &  0.95 c  &MF,LFM,SC &  7.92  &  7.80   & 0.31 &  7.74 & STM   \\  
 NGC55     &           &   9   &  1.45    &  G      &  9.47  &  8.99  & 0.38 &  8.34 &  RM     \\  
 IC10      &           &  10   &  0.83 c  &SHKD,WWRSH&  8.76  &  7.96  &1.08: &  8.18 & LPRSTP \\  
 SMC       &           &   9   &  0.06 c  & F       &  8.82  &  8.68  & 0.36 &  8.03 &  RM     \\  
 IC1613    & DDO 8     &  10   &  0.75 c  & MF,LFM  &  8.02  &  7.80  & 0.65 &  7.71 &  RM     \\  
 NGC1156   &           &  10   &  7.80    &  KMG    &  9.44  &  9.03  & 0.38 &  8.23 &  VSC    \\  
 NGC1569   & 7 Zw 16   &  10   &  2.01    &  KT     &  9.02  &  7.97  & 0.23 &  8.16 &  RM     \\  
 LMC       &           &   9   &  0.05    &  LFM    &  9.31  &  8.58  & 0.43 &  8.35 &  RM     \\  
 NGC2366   & DDO 42    &  10   &  3.44 c  &  TSHM   &  8.85  &  8.91  & 0.45 &  7.92 &  RM     \\  
 DDO47     &           &  10   &  4.27    &  GKT    &  8.05  &  8.44 & 0.45 &  7.85  &  SKH  \\  
 HolmbergII& DDO 50    &  10   &  3.05 c  &  HSD    &  8.80  &  8.82  & 0.39 &  7.92 &  RM   \\  
 DDO53     &           &   9   &  3.40    &  KT     &  7.49  &  7.81  & 0.31 &  7.62  &  SKH  \\  
 Leo A     & DDO 69    &  10   &  2.20 c  &  HSKD   &  7.80  &  7.87   & 0.26 &  7.30  &  SKH  \\  
 Sex B     & DDO 70    &  10   &  1.44 c  &  MF     &  7.88  &  7.65 & 0.47 &  8.11  &  MAM    \\  
 NGC3109   & DDO 236   &   9   &  1.36 c  &  MPC    &  8.66  &  8.74  &      &  8.06 &  RM     \\  
 Sex A     & DDO 75    &  10   &  1.45 c  &  SMF    &  7.82  &  7.94  & 0.35 &  7.49 &  SKH    \\  
 IC2574    & DDO 81    &   9   &  3.40    &  KT     &  9.12  &  9.05 & 0.34 &  8.08  &  RM     \\  
 NGC4214   &           &  10   &  4.10    & Letal   &  9.37  &  9.05  & 0.43 &  8.23 &  RM     \\  
 NGC4449   &           &  10   &  5.40    &  HT     &  9.69  &  9.47  & 0.37 &  8.32 &  SKH    \\  
 GR8       & DDO 155   &  10   &  2.24 c  &  TSHD   &  7.04  &  7.00 & 0.30 &  7.68  &  MAM    \\  
 NGC5253   &        &  10   &  4.09 c  &  Setal  &  9.23  &  8.36  & 0.28 &  8.16 &PSTE, RSRWR \\  
 NGC5408   & Tol 116   &  10   &  3.24    &  RM     &  8.54  &   8.21 & 0.42 &  8.01      &  RM     \\  
 IC4662    &      &  10   &  2.00    & H-MMM   &  8.29  &  8.08  & 0.29 &  8.06      & H-MMM   \\  
 NGC6822   &      &  10   &  0.51 c  & MF,LFM  &  8.25  &  7.84  &      &  8.23      &  RM     \\  
 Pegasus   &  & 10  &  0.76    & Getal   &  6.77  &  6.51  & 0.54 &  7.93   &  SBK    \\  
\hline \hline
\end{tabular}
\end{flushleft}

\vspace{0.2cm}

Note to Table:  

      The cepheid distances are labeled with letter c.

       The B-V value for IC10 is taken from VSC.
                
        HI measurement for DDO53 is taken from vDetal
\vspace{0.2cm}

List of references to Table:
           F --     Freedman, 1988; 
           G --     Graham, 1982;
           Getal -- Gallagher et al, 1998;
           GKT --   Georgiev, Karachentsev, Tikhonov, 1997;
           H-MMM -- Heydari-Malayeri, Melnick, Martin 1990;
           HSD --   Hoessel, Saha, Danielson, 1998;
           HSKD --  Hoessel, Saha, Krist, Danielson, 1994;
           HT --    Hunter, Thronson, 1996;  
           KMG --   Karachentsev, Musella, Grimaldi, 1996;
           KSRWR -- Kobulnicky, Skillman, Roy, Walsh, Rosa, 1997;
           KT --    Karachentsev, Tikhonov, 1994;
           Letal -- Leitherer, Vacca, Conti, Filippenko, Robert, Sargent, 1996;
           LFM --   Lee, Freedman, Madore, 1993;
           LPRSTP --Lequeux, Peimbert, Rayo, Serrano, Torres-Peimbert, 1979;
           MAM --   Moles, Aparicio, Masegosa, 1990;
           MF --    Madore, Freedman, 1991;
           MPC --   Musella, Piotto, Capaccioli, 1997;
           PSTE --  Pagel, Simonson, Terlevich, Edmunds, 1992;
           RM --    Richer, McCall, 1995;
           SC --    Sandage, Carlson, 1985;
           SBK --   Skillman, Bomans, Kobulnicky, 1997;
           Setal -- Sandage, Saha, Tammann, Labhard, Schwengeler, Panagia, Macchetto, 1994; 
           SHKD --  Saha, Hoessel, Krist, Danielson, 1996;
           SKH --   Skillman, Kennicutt, Hodge, 1989;
           SMF --   Sakai, Madore, Freedman, 1996;
           STM --   Skillman, Terlevich, Melnick, 1989;
           TSHD --  Tolstoy, Saha, Hoessel, Danielson, 1995; 
           TSHM --  Tolstoy, Saha, Hoessel, McQuade, 1995; 
           vDetal --  van Driel, Kraan-Korteweg, Binggeli, Huchtmeier, 1998;
           VSC --   Vigroux, Stasinska, Comte, 1987;
           WWRSH -- Wilson, Welch, Reid, Saha, Hoessel, 1996 
\end{table*}

In our previous work (Pilyugin \& Ferrini 1998), we defined the oxygen abundance 
deficiency in galaxies as 1 minus the ratio of the observed  oxygen abundance 
to that predicted by the closed-box model for the same gas mass fraction. 
Variations of the oxygen abundance deficiency from galaxy to galaxy can only 
be caused by differences in gas exchange between galaxies  and their environments.
In turn, the oxygen abundance deficiency can be considered as an indicator 
of mass exchange, in particular by galactic winds.
The goal of the present study is to study this problem quantitatively.
In Section 2 we present the observational data base.
The stellar mass to luminosity ratio in irregular galaxies is derived in 
Section 3, the values of the oxygen abundance deficiency for a number of
irregulars are obtained in Section 4. Section 5 contains discussion and a summary.

\section{Observational data}

The data base used in this study consists of published B magnitudes of
irregular galaxies, B--V colors, distances, oxygen abundances, HI emission fluxes, 
and radial distributions of HI. The variations in oxygen abundance from region 
to region in the best studied irregular galaxies are always within the observational 
errors (Pagel et al 1978, Skillman et al 1989) and there is no detectable radial 
oxygen abundance gradient. Then, the chemical composition of irregular galaxy can
be specified by a single parameter.

Our sample contains 25 irregular galaxies for which we have collected the 
relevant observational data, given with references in Table \ref{table:obsdat}.
The commonly used name(s) are given in columns 1 and 2, the Hubble 
type (T from  de Vaucouleurs et al 1991, RC3) in column 3, the adopted distance 
in column 4 (references in column 5). The cepheid distances are labeled with letter c.
The B luminosity computed from the adopted distance and the $B^{T}_{0}$  magnitude 
from  RC3 is given in column 6, and the mass of atomic hydrogen from the 
$m_{21}$  magnitude from RC3 in column 7. The B-V color  from RC3 is given in 
column 8 and the oxygen abundance  in column 9 (references in column 10).

\section{Stellar mass to luminosity ratio}

Irregular galaxies can contain a large amount of dark matter  (Carignan \& 
Beaulieu 1989, Kumai \& Tosa 1992) which might not take part directly to the chemical 
evolution. The luminous mass (mass of gas, stars, and stellar remnants) should be used 
rather than the dynamical mass in investigations of the chemical properties of irregular 
galaxies. The mass of stars $M_{s}$ including stellar remnants can be derived from 
the luminosity of the galaxy if the stellar mass to light ratio $\Upsilon_{s}=M_{s}/L_{B}$
can be determined. It depends on the star formation history (SFH) in the galaxy. 
For a few nearby irregular galaxies the star formation history can
be derived from the color-magnitude diagram of a large number of individual 
stars. Such determinations exist for the LMC (Butcher 1977,
Frogel 1984, Bertelli et al 1992, Elson et al 1994),  NGC~6822 (Gallart et
al 1996a,b), the Pegasus dwarf (Aparicio et al 1997, Gallagher et al 1998), 
and Sextans A (Dohm-Palmer et al 1997). 
Two variants of star formation histories will be used for Pegasus dwarf
according to two different estimates of the distance: 0.95 Mpc from Aparicio et al (1997), 
and 0.76 Mpc from Gallagher et al (1998).

%++++++++++++++++++++++++++++++++++++++++++++++++++++++ Table color
\begin{table}
\caption[]{\label{table:color}
Observed (U-B)$_{O}$, (B-V)$_{O}$, computed (U-B)$_{C}$, (B-V)$_{C}$  
colors, and computed stellar mass to luminosity ratio $M_{s}/L_{B}$ for 
irregular galaxies with known star formation histories. The observed colors 
are taken from RC3 for the LMC and the Pegasus dwarf, and from Hunter \& Plummer (1996)
for Sextans A.}
\begin{flushleft}
\begin{tabular}{lrrccc} \hline 
Galaxy        & (U-B)$_{O}$ & (U-B)$_{C}$ & (B-V)$_{O}$ & (B-V)$_{C}$ & $M_{s}/L_{B}$  \\ \hline
LMC           &  -- 0.06    &  -- 0.08    &  0.43       &  0.44       &   0.47     \\  
NGC6822       &             &  -- 0.37    &             &  0.24       &   0.35     \\  
Pegasus$^{1}$ &     0.06    &  -- 0.07    &  0.54       &  0.48       &   0.75     \\  
Pegasus$^{2}$ &     0.06    &     0.02    &  0.54       &  0.59       &   0.94     \\  
Sextans A     &  -- 0.36    &  -- 0.38    &  0.26       &  0.24       &   0.29     \\  \hline 
\end{tabular}
\end{flushleft}

\vspace{0.2cm}

Notes to Table:
           1    -- SFH from Aparicio et al, 1997;
           2    -- SFH from Gallagher et al, 1998

\end{table}

Using our model for the chemical and photometric evolution of an one-zone system  
(Pilyugin \& Ferrini 1998), the mass to luminosity ratios $M_{s}/L_{B}$ and 
colors U-B and B-V have been computed for galaxies with known star 
formation histories. The case S of nucleosynthesis has been 
adopted (Pilyugin \& Ferrini 1998). The computed and observed U-B and B-V 
colors as well as the computed star mass to luminosity ratios of these galaxies are 
given in Table \ref{table:color}. The observed colors of galaxies were taken from RC3
if not indicated otherwise.

\begin{figure}[thb]
\vspace{7.5cm}
\includegraphics{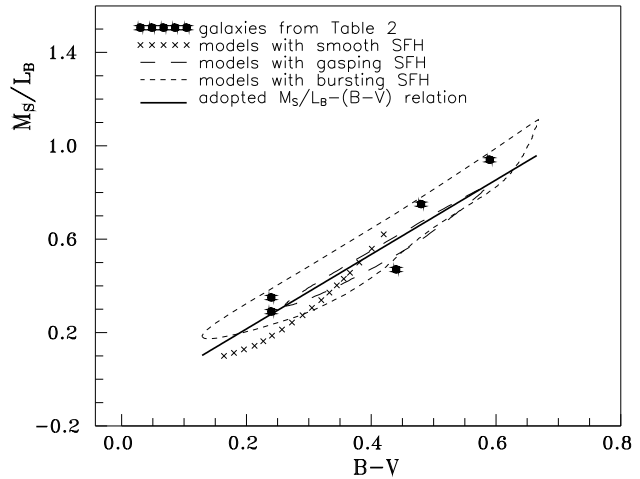}
\caption{\label{figure:9368f1}
The positions of galaxies with known star formation history (SFH) (filled circles)
in the diagram stellar mass to luminosity ratio $M_{s}/L_{B}$ versus B-V.
These positions are compared with the results of models with 
smooth SFHs (crosses) for different values of $\tau$, and of models with gasping
(long-dashed line) and bursting (short-dashed line) SFHs for different values 
of $t_{0}$. The solid line is the $M_{s}/L_{B}$ -- (B-V) relationship adopted in this paper.}
\end{figure}

The computed positions of galaxies from Table \ref{table:color} in the 
$M_{s}/L_{B}$ - (B-V) diagram are shown in  Fig.\ref{figure:9368f1}.

%============================================================Fig 9368f1 ml-bv
 
The Hubble sequence has been related to star formation histories (see e.g. Sandage 1986); 
in particular it is expected that the average star 
formation rate in irregular galaxy increases slightly with time. 
Since we are not interested here in the comprehension of the physics of the 
evolution of irregulars, we adopt simple analytical expressions for the star 
formation rate, depending on a time scale $\tau$:

\begin{equation}
\psi (t) \propto     \exp(t/\tau) .
\end{equation}

\begin{figure}[thb]
\vspace{7.5cm}
\includegraphics{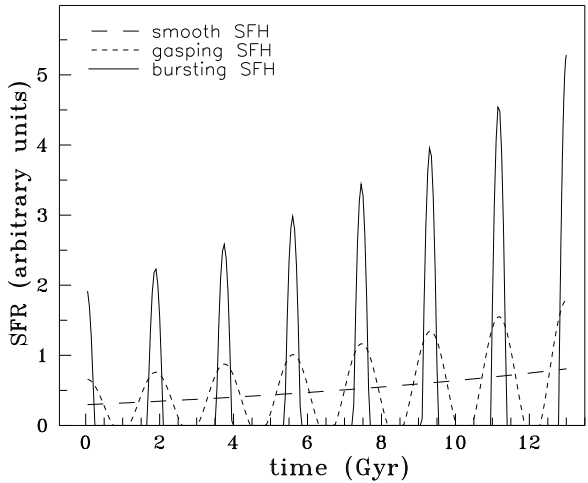}
\caption{\label{figure:9368f2}
Continuous (long-dashed line), bursting (solid line), and
 gasping (short-dashed line) star formation histories (SFH).}
\end{figure}

\begin{figure}[thb]
\vspace{7.5cm}
\includegraphics{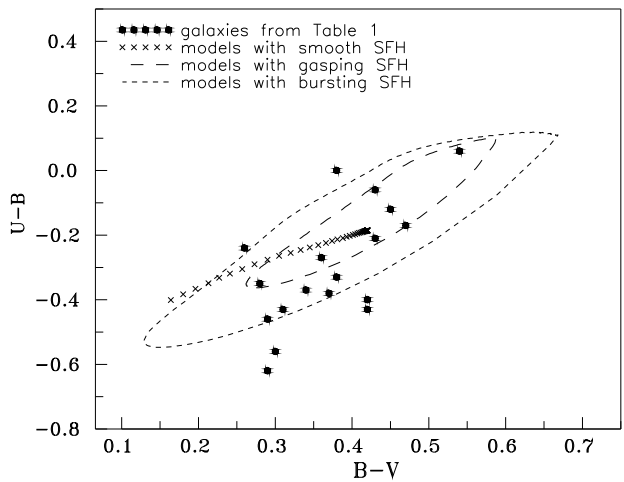}
 
\caption{\label{figure:9368f3}
The positions of galaxies (filled circles) in the U-B, B-V two color diagram.
They are compared with the results of models with 
continuous SFHs (crosses) for different values of $\tau$, models with gasping
(long-dashed line) and bursting (short-dashed line) SFHs for different values 
of $t_{0}$.}
\end{figure}

A grid of models with smooth star formation history for different values of 
$\tau$ from 1 Gyr (a strongly increasing star formation rate) 
to 500 Gyr (an almost constant star formation rate) have been computed. 
The star formation history with $\tau$=13 Gyr is shown in the Fig.\ref{figure:9368f2} by 
a long-dashed line. The positions of these models in the $M_{s}/L_{B}$ versus B-V 
diagram are shown in the Fig.\ref{figure:9368f1} by crosses. The agreement 
between these models and the representative points of galaxies is not good.
In particular, the values of B-V larger than $\sim$ 0.4 observed in
some irregulars cannot be reproduced.
The predictions of these models together with positions of galaxies from
Table \ref{table:obsdat} are shown on Fig.\ref{figure:9368f3}
in the U-B, B-V diagram. It is clear that 
models with smooth star formation histories cannot reproduce
the observed positions of irregular galaxies in this diagram.

%============================================================Fig 9368f2 modsfh.
 
%============================================================Fig 9368f3 ub-bv

This disagreement is a consequence of the hypothesis that the SFH is smooth. 
Actually star formation in irregular galaxies is not snooth: active periods 
alternate with quiet periods. In order to take this into 
account, models with non--smooth star formation histories have been computed. 
We describe it by the expression

\begin{equation}
\psi (t) \propto  \left\{ \begin{array}{ll}
     A(t) \; \exp{(t/\tau)} & \,\, {\rm if} \;\;\;\; A(t) > 0 , \\
          0   & \,\, {\rm if} \;\;\;\; A(t) \leq  0 ,
        \end{array}
         \right.
\end{equation}
where
\begin{equation}
     A(t) = cos (t/\tau _{P} + t_{0}) + c .
\end{equation}
The parameter $\tau _{P}$, which  indicates the number of cycles of 
star formation during the lifetime of a galaxy $t_{gal}$  
(that we adopted to be $t_{gal}$ = 13Gyr), is defined as

\begin{equation}
     t_{gal} / \tau _{P} = 2 n \pi  .
\end{equation}

\begin{figure}[thb]
\vspace{13.8cm}
\includegraphics{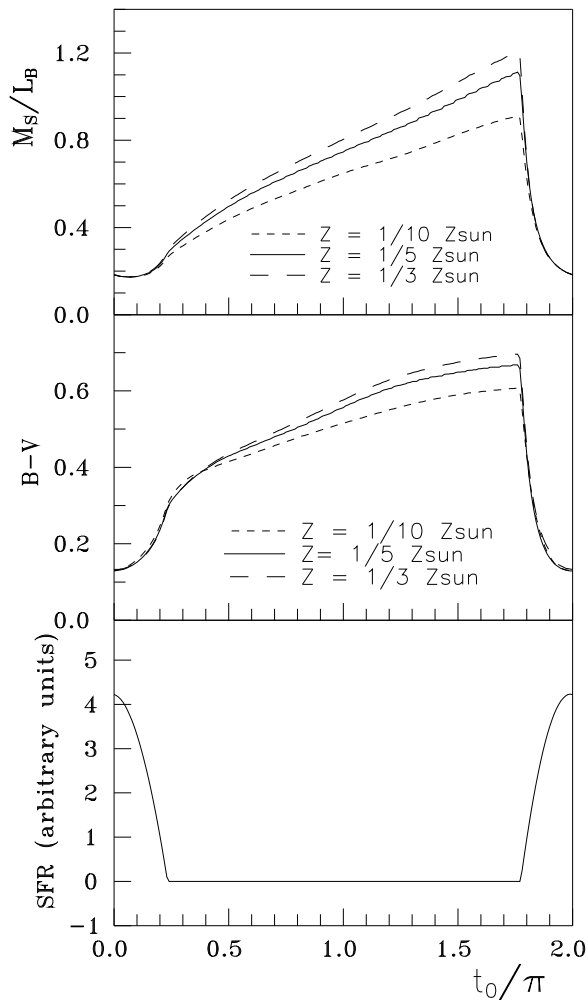}
\caption{\label{figure:9368f4}
The present--time star formation rate, the B-V color, and the $M_{s}/L_{B}$ ratio
as functions of $t_{0}$ for models with bursting star 
formation history.}
\end{figure}

Each cycle consists of an epoch with active star formation and an epoch without 
star formation.  Their relative duration is governed by the parameter $c$. The case $c=0$ 
corresponds to a periodic star formation history, in which the durations are equal.
The values $-1 < c < 0$ corresponds to shorter epochs with active star formation.
This case will be referred to as a ``bursting'' star formation history. 
The case with $c$ = --0.75, $\tau$ = 13 Gyr, $t_{0}$ = 0, and n = 7  
is shown in Fig.\ref{figure:9368f2}. The values $0<c<1$ corresponds to the opposite case.
Following Tosi (1993), this case will be referred to as a ``gasping'' star 
formation history. The case with $c$ = 0.75, $\tau$ = 13 Gyr, $t_{0}$ = 0, and n = 7  is 
shown in  Fig.\ref{figure:9368f2}. The point in the cycle of star formation at the present
time is governed by the choice of $t_{0}$ (Fig.\ref{figure:9368f4}). 
As it can be seen in Fig.\ref{figure:9368f4}, the B-V color and stellar 
mass to luminosity ratio $M_{s}/L_{B}$ are strongly dependent on $t_{0}$
(Fig.\ref{figure:9368f4} shows the evolutionary change of B-V and 
$M_{s}/L_{B}$ during the last cycle of star formation). The values of B-V and 
$M_{s}/L_{B}$ have a maximum at the point just before
the beginning of the star formation phase.
The values B-V and $M_{s}/L_{B}$ decrease with increasing 
of star formation rate, pass through a minimum at the maximum  star 
formation rate, and then increase.

%============================================================Fig 9368f4 to-par.

A grid of models with gasping star formation history has been computed. 
The positions of these models in the $M_{s}/L_{B}$ versus (B-V) diagram 
for various values of $t_{0}$ and for fixed values of other parameters 
($\tau$=13 Gyr, n=7, $c$=0.75)  are shown in Fig.\ref{figure:9368f1}. 
A similar grid of models with bursting star formation history 
has also be computed with $\tau$=13 Gyr, n=7, $c$=--0.75
(Fig.\ref{figure:9368f1}). The fit with observations is now good.
It is also good in the U-B versus B-V diagram. Some remaining 
discrepancies might be explained by the contribution of the ionized gas emission to 
the radiation of galaxy, which is not taken into account in the present models.
Fig. 1 and 3 suggest that the mass to luminosity ratio for and irregular galaxy with 
unknown star formation history can be estimated from its measured B-V color and 
relationship between stellar mass to luminosity ratio and B-V color.

\begin{figure}[thb]
\vspace{7.5cm}
\includegraphics{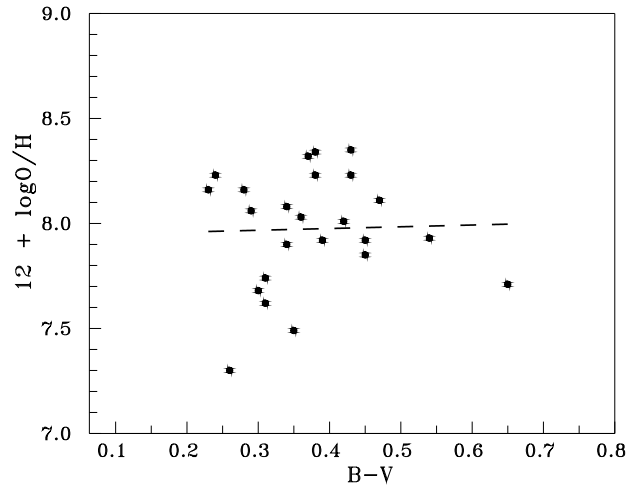}
\caption{\label{figure:9368f5}
The positions of irregular galaxies from our sample in the B-V versus O/H
diagram. The line is the formal best fit to the data. 
There is no correlation between the B-V color and the present-day oxygen 
abundance.}
\end{figure}

Up to here we have only considered models with a fixed present oxygen abundance 
12+log O/H=8.2 ($O/H \approx 1/5(O/H)_{\odot}$).
The mass to luminosity ratio depends not only on the star
formation history but also on the metallicity. To examine
the dependence of the mass to luminosity ratio on the metallicity, models with
bursting star formation history and with present oxygen abundances  
 12+log O/H=7.9 ($O/H \approx 1/10(O/H)_{\odot}$) and
12+log O/H=8.4 ($O/H \approx 1/3(O/H)_{\odot}$) have been computed.
For the range of observed B-V (0.3 to 0.5) the differences in $M_{s}/L_{B}$ predicted
by models with 12+log O/H=7.9 and 12+log O/H=8.4 with those predicted by models with
12+log O/H=8.2 are smaller than 15 per cent (Fig.\ref{figure:9368f4}). 
On the other hand, Fig.\ref{figure:9368f5} shows that there is no correlation
between B-V and the present-day oxygen abundance.
This shows that B-V is mainly determined by the star formation history in the 
recent past. Therefore the  relationship between 
stellar mass to luminosity ratio $M_{s}/L_{B}$ and B-V color can be considered as 
independent on the present oxygen abundance.

%============================================================Fig 9368f5 bv-oh

We adopt as a first approximation a linear relationship between 
mass to luminosity ratio $M_{s}/L_{B}$ and B-V color (Fig.1)
\begin{equation}
     \Upsilon_{s} = -0.10 \pm 0.15 + 1.60 \, (B-V).  \label{eq:mlberr}  
\end{equation}
The error on the intercept includes the range of models.

\section{Oxygen abundance deficiency in irregular galaxies}

\subsection{Hydrogen content in irregular galaxies}

The size of irregular galaxies in HI is larger than their optical size (Salpeter
\& Hoffman 1996, and references therein).
As we have seen, the variations of oxygen abundance from region to region in the best
studied irregular galaxies are within observational errors, a strong evidence that 
irregular galaxies are well mixed system inside the optical diameter.
If the zone of the mixing extends to the outer HI then the gas outside the optical diameter 
can contain a significant part of the heavy elements produced by the stars and,
therefore, can play an important role in the chemical evolution of the galaxy. 
Hunter \& Gallagher (1985) have found that the presence or absence of 
extended HI envelopes does not appear to be reflected in the properties of the 
optical components of irregular galaxies. They concluded that only the gas 
within the optical limit plays an important role in the life of a galaxy. 

%++++++++++++++++++++++++++++++++++++++++++++++++++++++ Table fopt
\begin{table}
\caption[]{\label{table:fopt} 
Fraction of atomic hydrogen mass within the optical diameter $f_{opt}$,
hydrogen to optical radius ratio $R_{H}/R_{0}$, with references,
for a number of irregular galaxies in the present sample}
%\begin{flushleft}
\begin{center}
\begin{tabular}{lccl} \hline 
%              &           &               &       \\  
galaxy name   & $f_{opt}$ & $R_{H}/R_{25}$ &  ref  \\   \hline 
NGC 55        &   0.92    &    1.14       &  PCW  \\  
IC 1613       &   0.68    &    1.41       &  LS   \\  
NGC 1156      &   0.84    &    1.46       &  BW   \\  
NGC 2366      &   0.67    &    1.65       &  WKA  \\  
Holmberg II   &   0.56    &    1.79       &  PWBR \\ 
Leo A         &   0.51    &    1.84       &  YL   \\
NGC 3109      &   0.61    &    1.50       &  JC   \\    
Sextans A     &   0.48    &    1.77       &  STTW \\  
IC 2574       &   0.64    &    1.40       &  MCR  \\  
GR 8          &   0.36    &    2.02       &  CBF  \\  \hline 
\end{tabular}
%\end{flushleft}
\end{center}

\vspace{0.2cm}

List of references to Table:
           BW --    Broeils, van Woerden, 1994;
           CBF --   Carignan, Beaulieu, Freeman, 1990;
           JC --    Jobin, Carignan, 1990;
           LS --    Lake, Skillman, 1989;
           MCR --   Martimbeau, Carignan, Roy, 1994;
           PCW --   Puche, Carignan, Wainscoat, 1991;
           PWBR --  Puche, Westpfahl, Brinks, Roy, 1992;
           STTW --  Skillman, Terlevich, Teuben, van Woerden, 1988;
           WKA --   Wevers, van der Kruit, Allen, 1986;
           YL --    Young, Lo, 1996 
\end{table}

The mass of atomic hydrogen in the irregular galaxies listed in Table 
\ref{table:obsdat} is the total mass.
For a number of irregular galaxies from our sample there are
measurements of radial distribution of surface mass density of 
atomic hydrogen that allows to determine the fraction of HI mass 
within the optical diameter $f_{opt}$. Table \ref{table:fopt} gives the fraction 
$f_{opt}$ and the ratio of the hydrogen radius $R_H$ (radius at face-on surface 
mass density of atomic 
hydrogen of 1 $M_{\odot}\,pc^{-2}$) to the optical radius $R_{25}$ (from RC3).
Table \ref{table:fopt} shows the 
fraction of HI mass within the optical diameter of a galaxy varies widely
from $f_{opt}$ = 0.36 for GR 8 to $f_{opt}$ = 0.92 for NGC 55.

\subsection{Oxygen abundance deficiency in irregular galaxies}

The gas mass fraction $\mu$ is defined as 
\begin{equation}
\mu = M_{g}/(M_{g}+M_{s})  ,
\end{equation}
where $M_{g}$ is the mass of gas within the optical radius $R_{25}$ of the galaxy.
The mass of stars is derived from the blue luminosity and the mass to luminosity ratio
$\Upsilon_{s}$
\begin{equation}
     M_{s} = \Upsilon_{s} \times L_{B} .
\end{equation}
The values of $\Upsilon_{s}$ computed using relation (\ref{eq:mlberr}) are
listed in column 2 of Table \ref{table:comdat}.
The mass of gas within the optical edge of a galaxy is taken as 
\begin{equation}
     M_{gas} =f_{opt} \times (1.4 \times  (1 + r_{H_{2}}) \times M_{HI})
\end{equation}
where the factors $r_{H_{2}}$ and 1.4 are introduced to take the contributions of
molecular hydrogen and helium to gas mass into account, respectively.
$f_{opt}$ comes from  Table \ref{table:fopt}.
The correction for the presence of molecular hydrogen is uncertain
since the $H_{2}$ content has been determined from CO line observations
 in a few irregular galaxies only.
Fortunately, the average global $H_{2}$ to HI mass ratio $r_{H_{2}} = H_{2}/HI$
is not large for these galaxies, $r_{H_{2}}$ = 0.2$\pm$0.04,
(Israel 1997). Here we adopt $r_{H_{2}}$ = 0.2 for all the galaxies.
The values of $\mu$ are given in column 3 of Table \ref{table:comdat}. 

%++++++++++++++++++++++++++++++++++++++++++++++++++++++ Table comdat
\begin{table}
\caption[]{\label{table:comdat}
Computed characteristics (star mass to luminosity ratio $\Upsilon_{s} = M_{s}/L_{B}$, gas 
mass fraction $\mu$, oxygen abundance deficiency  $\eta$, and total luminous 
mass to luminosity ratio $\Upsilon_{t}=M_{t}/L_{B}$) for the irregular galaxies of the present 
sample}
\begin{center}
\begin{tabular}{lcllcccc} \hline 
           &           &\multicolumn{3}{c|}{within $R_{25}$}  & \multicolumn{3}{c}{within $R_{H}$}   \\  \cline{3-8}
name       & $\Upsilon_{s}$ &\multicolumn{1}{c}{ $\mu$} & $\eta$   
                                    &\multicolumn{1}{c|}{ $\Upsilon_{t}$ }   
                  & $\mu$   & $\eta$  & $\Upsilon_{t}$  \\  \hline
 WLM       &  0.40     &         &           &          &  0.76   & 0.71    &  1.67          \\
 NGC55     &  0.51     &  0.50   &  0.55     &  1.02    &  0.52   & 0.52    &  1.06          \\
 SMC       &  0.48     &         &           &          &  0.72   & 0.54    &  1.69          \\
 IC1613    &  0.94     &  0.42   &  0.92     &  1.63    &  0.52   & 0.89    &  1.95          \\
 NGC1156   &  0.51     &  0.52   &  0.64     &  1.06    &  0.56   & 0.58    &  1.16          \\
 NGC1569   &  0.27     &         &           &          &  0.36   & 0.80    &  0.42          \\
 LMC       &  0.47     &         &           &          &  0.40   & 0.66    &  0.78          \\
 NGC2366   &  0.62     &  0.68   &  0.70     &  1.91    &  0.76   & 0.57    &  2.55          \\
 DDO47     &  0.62     &         &           &          &  0.87   & 0.26    &  4.74          \\
 HolmbII   &  0.52     &  0.65   &  0.72     &  1.51    &  0.77   & 0.54    &  2.28          \\
 DDO53     &  0.40     &         &           &          &  0.90   & 0.43    &  3.91          \\
 Leo A     &  0.32     &  0.76   &  0.90     &  1.32    &  0.86   & 0.80    &  2.29          \\
 Sex B     &  0.65     &         &           &          &  0.60   & 0.64    &  1.64          \\
 Sex A     &  0.29     &  0.79   &  0.82     &  1.35    &  0.88   & 0.63    &  2.50          \\
 IC2574    &  0.44     &  0.67   &  0.57     &  1.36    &  0.76   & 0.36    &  1.87          \\
 NGC4214   &  0.59     &         &           &          &  0.58   & 0.56    &  1.39          \\
 NGC4449   &  0.49     &         &           &          &  0.67   & 0.25    &  1.50          \\
 GR8       &  0.38     &  0.59   &  0.87     &  0.93    &  0.80   & 0.69    &  1.91          \\
 NGC5253   &  0.35     &         &           &          &  0.39   & 0.78    &  0.57          \\
 NGC5408   &  0.57     &         &           &          &  0.58   & 0.74    &  1.36          \\
 IC4662    &  0.36     &         &           &          &  0.74   & 0.45    &  1.40          \\
 NGC6822   &  0.35     &         &           &          &  0.65   & 0.44    &  1.00          \\
 Pegasus   &  0.84     &         &           &          &  0.52   & 0.81    &  1.76          \\  \hline
\end{tabular}
\end{center}
\end{table}

%============================================================Fig 9368f6 mu-oh-ro
\begin{figure}[thb]
\vspace{7.5cm}
\includegraphics{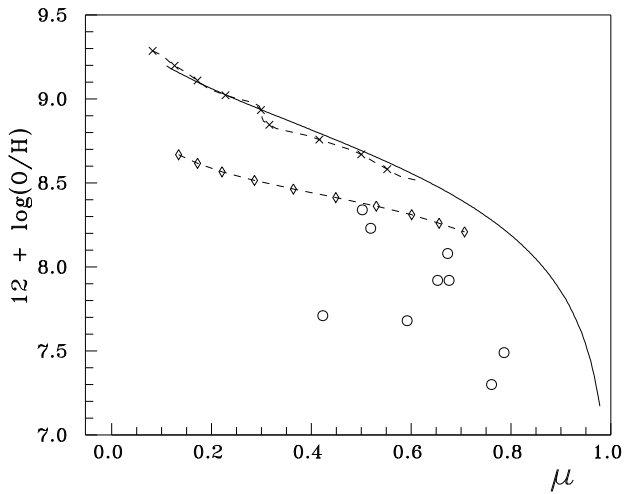}
\caption{\label{figure:9368f6}
The positions of irregular galaxies (open circles) in a diagram O/H vs. the gas
fraction within the optical radius.
The results of one-zone closed-box models with different present-day gas 
mass fraction (full curve) is presented by the solid line. The 
positions of two spiral galaxies NGC~5457 (crosses) and 
NGC~2403 (lozenges) are also shown for comparison. 
The data for these spirals are from Pilyugin \& Ferrini 1998.
}
\end{figure}
 
The global oxygen abundance deficiency is defined as 
$\eta = 1 - Z_{O}^{obs}/Z_{O}^{CB}$. The values of $\eta$ are given in column 
4 of Table \ref{table:comdat}. The positions of irregular galaxies in the 
$\mu$ -- O/H diagram shown on  Fig.\ref{figure:9368f6} by circles. The curve defined by the 
one-zone closed-box models with different gas mass fraction (this curve will 
be referred to as standard curve) is shown on Fig.\ref{figure:9368f6} as a 
solid line. For comparison we show the positions for various regions of the
 two spiral galaxies NGC~5457 
(dashed line with crosses) and NGC~2403 (dashed line with lozenges) 
(data from Pilyugin \& Ferrini 1998).
NGC~5457 has no oxygen abundance deficiency, and NGC~2403 has the largest
global oxygen abundance deficiency ($\eta = 0.57$) among the spiral galaxies
considered by Pilyugin and Ferrini (1998). As it can be seen in  
Fig.\ref{figure:9368f6}, all the irregular galaxies  are significantly 
displaced from the standard curve. Their values of oxygen abundance deficiency 
are also in excess of that in the dwarf spiral galaxy NGC~2403. The values 
of oxygen abundance deficiency range from $\sim$ 0.5 to $\sim$ 0.9. 
The relationships between the value of oxygen abundance deficiency and global 
parameters will be given in one of the next papers of this series.

Since irregular galaxies are gas-rich systems (as it can be seen from 
Table \ref{table:comdat} the mass of gas component exceeds the mass of the
stellar component in the galaxy) the total luminous mass to luminosity 
ratio $\Upsilon_{t}=M_{t}/L_{B} = (M_{s}+M_{gas})/L_{B}$ is significantly larger 
than  the value of star mass to luminosity ratio $\Upsilon_{s}$
(compare the data from columns 2 and 5 of Table \ref{table:comdat}).

Can the oxygen abundance deficiency within optical edge of irregular galaxy
be explained by the mixing of interstellar matter with gas outside the optical 
edge? To study this possibility, the oxygen abundance deficiency have
been re--computed under the assumption that all the gas associated with a galaxy
is well mixed. In this case the mass of gas in the galaxy is 
\begin{equation}
     M_{gas} =1.4 \times  (1 + r_{H_{2}}) \times M_{HI} .
\end{equation}
All the other values are derived in the same way as in the previous case.
 The values of $\mu$, $\eta$, and $M_{t}/L_{B}$ are listed in columns 6-8 of 
Table \ref{table:comdat}. 

%============================================================Fig 9368f7 mu-oh-rh
\begin{figure}[thb]
\vspace{7.5cm}
\includegraphics{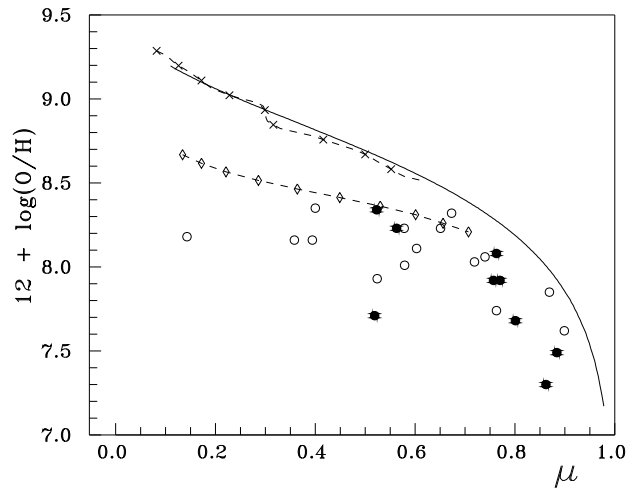}
\caption{\label{figure:9368f7}
Same as Fig. 6, but all the gas of the galaxy is now taken into consideration.
The positions of irregular galaxies with known fraction of gas mass within the 
optical boundaries $f_{opt}$ are shown 
by filled circles, and the positions of irregular galaxies with unknown $f_{opt}$ 
are shown by open circles.
The results of one-zone closed-box models with different present-day gas 
mass fraction (full curve) is presented by the solid line. The 
positions of two spiral galaxies NGC~5457 (crosses) and 
NGC~2403 (lozenges) are also shown for comparison. 
The data for these spirals are taken from Pilyugin \& Ferrini 1998.
}
\end{figure}
 
The positions of irregular galaxies in the $\mu$ -- O/H diagram in the present case 
are shown on  Fig.\ref{figure:9368f7}: significant displacements of irregular galaxies 
relative to the standard curve remain.  The
inclusion of extended HI envelope outside the optical radius 
only slightly decreases the displacement of irregular galaxies 
relative to the standard curve. 

\section{Conclusions}

We have found that all the irregular galaxies considered here have a significant 
oxygen abundance deficiency.  This is a strong 
argument in favor of an important role of mass exchange with 
the environment in the chemical evolution of these galaxies.
These conclusions remain even if one takes into consideration the gas envelope
observed outside the optical edge. This confirm the widespread idea that
a significant part of the heavy elements is ejected by irregular galaxies
in the intergalactic medium.

We emphasize that the relation between oxygen abundance vs. gas mass fraction
in the closed box model used in present study was
constructed so that the positions  of the most metal-rich
giant spiral galaxies (including NGC~5457) in the $\mu$ -O/H diagram are fitted 
by this relation (Pilyugin \& Ferrini 1998). Therefore the 
oxygen abundance deficiency of irregular galaxies can  also be 
considered as relative to the giant spiral galaxy NGC~5457.
It has been found  by Pilyugin \& Ferrini (1998) that the precise choice of
the oxygen production by individual stars and of the parameters of the initial
mass function are not critical in this differential comparison.

\begin{acknowledgements}
We thank Dr. N.Bergvall for his constructive comments on the manuscript.
We thank the referee, Prof. J.Lequeux, for the suggestions and comments that 
improved sensibly the presentation.  
L.P. thanks the Staff of Department of Physics, Section of Astronomy 
(University of Pisa) for hospitality. This study was partly supported by the 
INTAS grant No 97-0033. 
\end{acknowledgements}

\end{document}